\documentclass[prb,twocolumn,floatfix,showpacs,superscriptaddress]{revtex4-1}
\usepackage{graphics}
\usepackage{epsfig}
\usepackage{times}
\usepackage{bm}
\usepackage{braket}
\usepackage{color}

\usepackage{amsmath}	
\begin{document}
\title{Field-induced incommensurate ordering in the Heisenberg chains coupled by Ising interaction:\\ Model for the ytterbium aluminum perovskite YbAlO$_3$}
\author{Cli\`o Efthimia Agrapidis}
\affiliation{Institute for Theoretical Solid State Physics, IFW Dresden, 01069 Dresden, Germany}
\author{Jeroen van den Brink}
\affiliation{Institute for Theoretical Solid State Physics, IFW Dresden, 01069 Dresden, Germany}
\affiliation{Department of Physics, Technical University Dresden, 01069 Dresden, Germany}
\affiliation{Department of Physics, Washington University, St. Louis, MO 63130, USA}
\author{Satoshi Nishimoto}
\affiliation{Institute for Theoretical Solid State Physics, IFW Dresden, 01069 Dresden, Germany}
\affiliation{Department of Physics, Technical University Dresden, 01069 Dresden, Germany}

\date{\today}

\begin{abstract}
We study isotropic antiferromagnetic Heisenberg chains coupled by antiferromagnetic Ising interaction as an effective spin model for the ytterbium aluminum perovskite YbAlO$_3$. Using the density-matrix renormalization group (DMRG) method we calculate the magnetization curve, local spin, central charge, and dynamical spin structure factors in the presence of magnetic field. From the fitting of the experimental magnetization curve, the effective intrachain and interchain couplings are estimated as $J=2.3$~K and $J_{\rm ic}=0.8$~K, respectively. We can quantitatively explain the experimental observations: (i) phase transition from antiferromagnetic to incommensurate order at field 0.35~T, and (ii) quantum critical behaviors at the saturation field of 1.21~T. Furthermore, the low-energy excitations in the experimental inelastic neutron scattering spectra can be well described by our DMRG results of the dynamical structure factors.
\end{abstract}

\maketitle

\section{Introduction}

In low dimensional strongly correlated electron systems, the effects of interactions and quantum fluctuations are maximized.
This situation provides us with an ideal playground to study quantum phase transitions. In fact, quasi-one-dimensional (quasi-1D) Heisenberg magnets reveal a variety of exotic phases~\cite{Vasiliev18}. 

Quantum phase transitions at zero temperature are characterized by a quantum critical point (QCP)~\cite{Coleman05,Sachdev11}, where the critical fluctuations are scale-invariant and the system belongs to a universality class characterized by critical exponents, independent of the microscopic details of the system~\cite{Moriya85,Binney92,Timusk99}.  Physical properties over a wide range of temperatures above a QCP can be influenced by  critical fluctuations and physical quantities such as magnetic susceptibility and specific heat obey simple scaling laws in the critical exponent.

In the pure 1D limit, the isotropic spin-$\frac{1}{2}$ Heisenberg model belongs to a universality class, the so-called Tomonaga-Luttinger liquid (TLL), exhibiting a gapless mode in the elementary excitations at zero temperature. This means that the 1D system is extremely fragile to external perturbations such as further interactions. Typically, when the 1D chains are coupled, they crosses over from the exotic TLL to more conventional 2D or 3D order. In some cases, weaker interchain couplings can drastically change the magnetic properties and yield the dominant contribution to physical quantities~\cite{saturation11}. Therefore, even in the case of small interchain coupling, an uncoupled 1D chain is not always a good approximation and dimensional crossovers are a challenging and open problem.

Very recently, TLL behavior was reported in the ytterbium aluminum perovskite YbAlO$_3$~\cite{Wu19}. Below $T_{\rm N}=0.88$~K the system is antiferromagnetically ordered, namely, in a N\'eel state. With applying external magnetic field, a N\'eel to incommensurate (IC) phase transition occurs at $H_{\rm c}=0.35$~T, and the spins fully saturate at $H_{\rm s}=1.13$~T. Scaling behaviors associated with the TLL universality class were experimentally observed as the magnetic susceptibility $dM/dH \propto \tilde{H}^{-0.51}\phi(T/\tilde{H}^{1.04})$ and the specific heat $[C(H)-C(H_{\rm s})]/T \propto \tilde{H}^{-0.5}\psi(T/\tilde{H})$ where $M$ is magnetic field, $M_{\rm s}$ is saturation field, and $T$ is temperature. Moreover, the spinon confinement-deconfinement transition was confirmed by the inelastic neutron scattering measurements, and it was theoretically explained within a single Heisenberg chain in the presence of staggered magnetic field~\cite{Wu19}.

In this paper, we propose isotropic antiferromagnetic (AFM) spin-$\frac{1}{2}$ Heisenberg chains coupled by antiferromagnetic Ising interaction as an effective spin model for YbAlO$_3$. The density-matrix renormalization group (DMRG) method is employed to study this spin model. From the fitting of experimental magnetization curve, the effective intrachain and interchain couplings are estimated as $J=2.3$~K and $J_{\rm ic}=0.8$~K, respectively. To see the stability of AFM/IC orders and to confirm a TLL criticality at the saturation field, we calculate local spin and central charge in the presence of magnetic field, respectively. Then, we can the quantitatively explain the experimental observations: (i) phase transition from AFM to IC order at field 0.35~T, and (ii) quantum critical behavior with a saturation field of 1.21~T. Furthermore, we find that the low-energy excitations in the experimental inelastic neutron scattering spectra can be well described by our DMRG results of the dynamical spin structure factors.

The paper is organized as follows: In Sec.~II our Hamiltonian for the effective spin model for YbAlO$_3$ is explained and the applied numerical method is described. In Sec.~III we present our numerical results and discuss the relevance for the experimental observations. Finally we give a conclusion in Sec. IV.

\begin{figure}[tbh]
\centering
\includegraphics[width=0.6\linewidth]{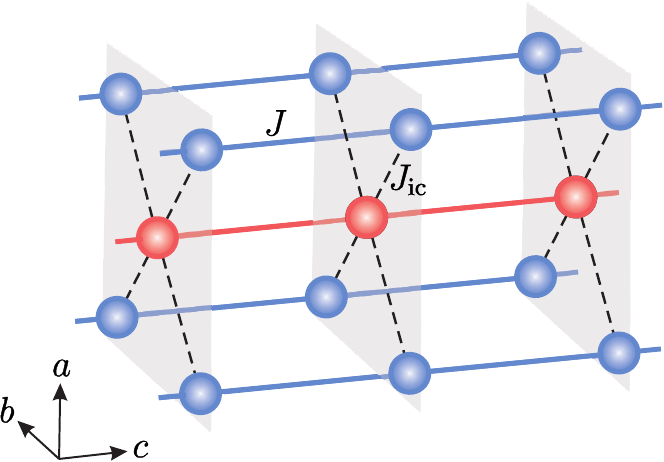}
\caption{
(a) Lattice structure of the effective spin model \eqref{ham1} for YbAlO$_3$. Red and blue chains denote two sublattices, where a bipartite square lattice is formed in the $a$-$b$ plane. Intrachain $J$ is isotropic Heisenberg and interchain $J_{\rm ic}$ is Ising interactions.
}
\label{lattice}
\end{figure}

\section{Model and method}

\subsection{Heisenberg chains coupled by Ising interaction}

We here propose an effective spin model to describe the fundamental magnetic properties of YbAlO$_3$. YbAlO$_3$ is insulating and the magnetic properties come from the Yb$^{3+}$ ions. The four Kramers doublets of the Yb$^{3+}$ ions are split by the crystal field and the doublet ground state $m_J=\pm\frac{7}{2}$ is well separated from the excited levels. Therefore, the low-energy magnetic properties can be described by a pseudo-spin-$\frac{1}{2}$ model~\cite{Radhakrishna81}. 
Magnetically, the system consists of two 1D sublattices (see Fig.~\ref{lattice}). The magnetic coupling between the two sublattices, i.e., interchain coupling, is similar to a dipole-dipole interaction and it leads to a Ising-type coupling $S^zS^z$. On the other hand, the intrachain coupling is dominantly created by an AFM super-exchange process. Furthermore, it is known that the exchange interaction of rare earth ions has the typical form $\vec{S}\cdot\vec{S}$~\cite{Wu16}. Thus, despite the highly anisotropic doublet state of Yb$^{3+}$ ions, one may assume an isotropic exchange coupling along the chain direction. The validity of this isotropic exchange coupling is also discussed in Appendix A. 

We  are thus dealing with isotropic Heisenberg chains coupled by Ising interaction as an effective spin model for YbAlO$_3$. The Hamiltonian is written as
\begin{align}
\nonumber
	H &= J \sum_i \sum_j \vec{S}_{i,j}\cdot\vec{S}_{i+1,j}+J_{\rm ic} \sum_i \sum_{j,j^\prime} S^z_{i,j} S^z_{i,j^\prime}\\
	&+H\sum_{i,j}S^z_{i,j},
	\label{ham1}
\end{align}
where $S^\alpha_{i,j}$ is the $\alpha$-component of spin-$\frac{1}{2}$ operator $\vec{S}_{i,j}$ at $i$-th site in $j$-th chain. The lattice structure is sketched in Fig.~\ref{lattice}. Since the interchain couplings form a bipartite square lattice in the $a$-$b$ plane and no exchange processes are allowed between the chains, the extension of the cluster in the $a$-$b$ plane can be reduced to a two-chain problem. Hence, in this paper, we study two isotropic Heisenberg chains coupled by Ising interaction. We note that, in order to count the strength of the interchain coupling consistently with the material, the interchain Ising coupling should be replaced by $\tilde{J}_{\rm ic}=4J_{\rm ic}/N_{\rm c}$, where $N_{\rm c}$ is the number of neighboring chains, i.e.,
$N_{\rm c}=1$ for two chains.

\subsection{Density-matrix renormalization group}

To examine the ground state of the system \eqref{ham1} we employ the DMRG technique which is a powerful numerical method for various (quasi) 1D quantum systems~\cite{White92}. For the calculation of static properties, we use the standard DMRG method. Either open or periodic boundary conditions are chosen dependent on the calculated quantity. We study clusters with length up to $L\times2=600\times2$ and keep up to $m=4000$ density-matrix eigenstates in the renormalization procedure. In this way, the maximum truncation error, i.e., the discarded weight, is less than $1 \times10^{-13}$. This high accuracy is naively expected because no quantum fluctuations are allowed between chains coupled by Ising interaction.

For the calculation of dynamical properties, we use the dynamical DMRG method which has been developed for calculating dynamical correlation functions at zero temperature in quantum lattice models~\cite{Jeckelmann02}. Since the dynamical DMRG algorithm performs best for open boundary conditions, we study a open cluster with length $L\times2=50\times2$. The dynamical DMRG approach is based on a variational principle so that we have to prepare a `good trial function' of the ground state with the density-matrix eigenstates. Therefore, we keep $m=1200$ to obtain the ground state in the first ten DMRG sweeps and keep $m=600$ to calculate the excitation spectrum. In this way, the maximum truncation error, i.e., the discarded weight, is about $1 \times10^{-5}$, while the maximum error in the ground-state and low-lying excited states energies is about $10^{-2}J$.


\section{Results and discussion}

\begin{figure}[tbh]
\centering
\includegraphics[width=0.9\linewidth]{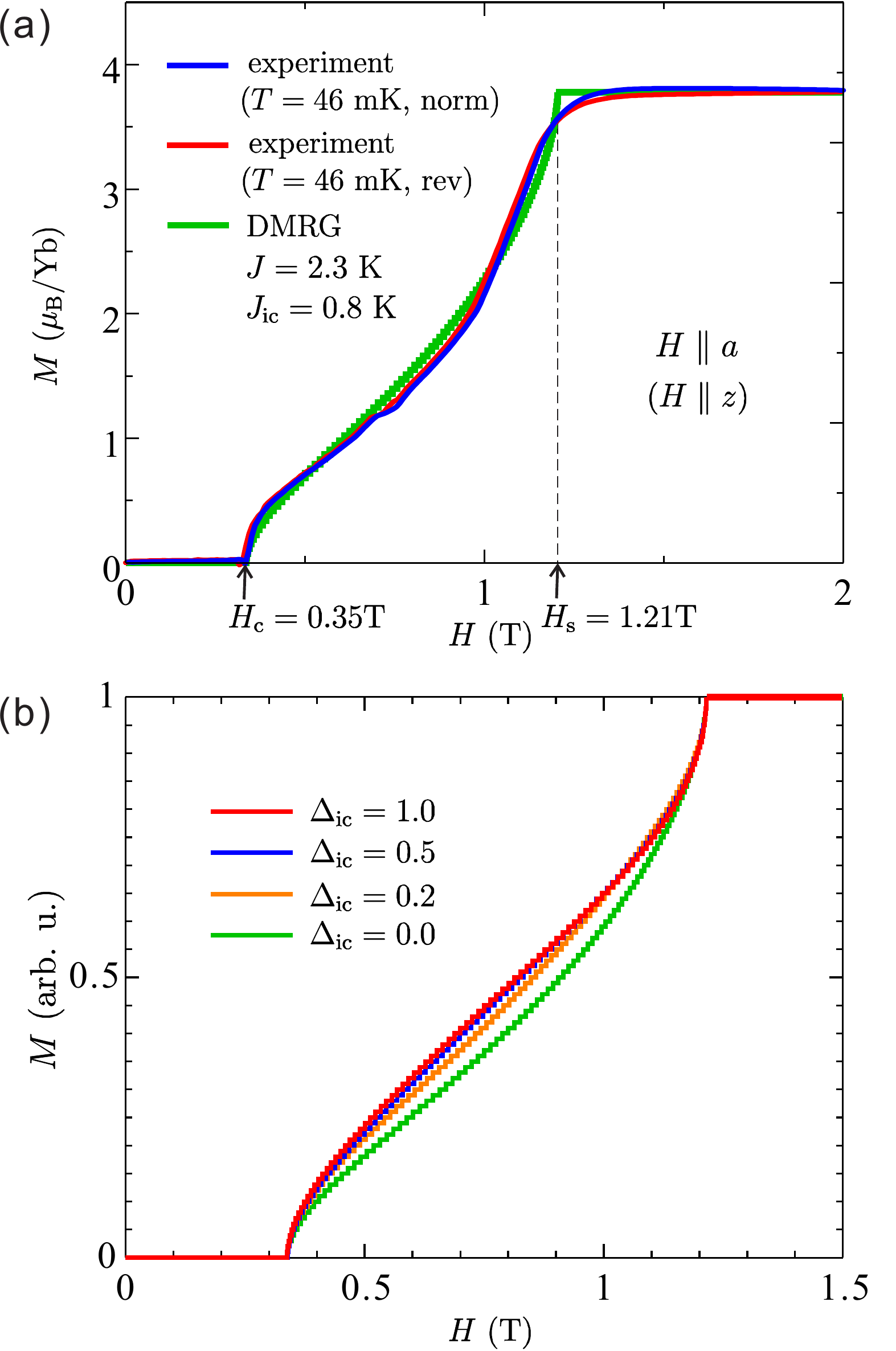}
\caption{
(a) Fitting of the low-temperature experimental magnetization curve for $H \parallel a$ by DMRG result with the system \eqref{ham1}, where $g^a(=g^z)=7.6$ is used. (b) DMRG results of magnetization curve for several kinds of interchain couplings with different XXZ anisotropy.
}
\label{magnetization}
\end{figure}

\subsection{Magnetization}

First, to estimate the effective coupling parameters for YbAlO$_3$, we perform a fitting of the experimental magnetization curve observed at $T=45$~mK under applied field $H$ parallel to the $a$-axis [see Fig.~\ref{magnetization}(a)]. The experimental magnetization curve exhibits; (i) an explicit gap between singlet and triplet states, the magnitude of which corresponds to the critical field $H_c=0.35$~T, and (ii) a sharp increase near the saturation $H_s=1.21$~T, which is a typical signature of strong quantum fluctuations in 1D Heisenberg systems. (Note that, although the saturation field has been estimated to be $H_{\rm s}=1.13$~T from the peak position of $dM/dH$ at $T=0.05$~K, it may be slightly shifted to higher field in the limit $T\to0$.) These two features can be reproduced by taking $J>0$ and $J_{\rm ic}>0$ in the system \eqref{ham1}: Due to the AFM Ising interchain coupling, a spinon propagation costs more at low field and quantum fluctuations are allowed only within each chain along the $c$-axis. This is consistent with the spinon confinement picture suggested in Ref.~\onlinecite{Wu19}.

At present, low-temperature magnetization measurement are available only for $H \parallel a$. As seen in Fig.~\ref{magnetization}(a), the observed saturation moment $M_{\rm s}=3.8 \mu_{\rm B}$/Yb leads to a large $g$-factor $g^a=7.6$ for $H \parallel a$. Since this value is much larger than those for the other directions $g^b \simeq g^c=0.46$, the magnetization is dominated by the $a$-component. Thus, the most reliable fitting  is chosen by using the magnetization curve for $H \parallel a$. The fitting result is given in Fig.~\ref{magnetization}(a), where a periodic cluster with $L\times2=100\times2$ sites is used for the DMRG calculations. The best fitting is obtained by assuming that the crystallographic $a$-axis is parallel with the $z$-axis ($z \parallel a$, $g^z=g^a$) in our model and by setting $J=2.3$K and $J_{\rm ic}=0.8$K. If we assume the interchain coupling to be ferromagnetic (FM), i.e., $J_{\rm ic}<0$, only a rough fitting is possible (see Appendix B). We note that the AFM and FM Ising interchain couplings give quantitatively the same results except for the anti-phase or in-phase chains at $M=0$; and, qualitatively similar results for $M>0$. Because of this fit to the experimental magnetization, we mostly consider the case of AFM Ising interchain coupling hereafter.

For confirmation, let us now examine the validity of the Ising-type interchain coupling. To test it, additional exchange terms $\Delta_{\rm ic}J_{\rm ic}(S^+_{i,j} S^-_{i,j^\prime}+S^-_{i,j} S^+_{i,j^\prime})$ are added to the Ising interchain coupling of the Hamiltonian~\eqref{ham1}. In Fig.~\ref{magnetization}(b) we show the magnetization curve for several values of $\Delta_{\rm ic}$, where the values of $J$ and $J_{\rm ic}$ are tuned to keep the ratio between $H_{\rm c}$ and $H_{\rm s}$. We find that the magnetization is lifted up from the Ising limit ($\Delta_{\rm ic}=0$) even by small $\Delta_{\rm ic}=0.2$ at the intermediate field $0.5~{\rm T} \lesssim H \lesssim 1~{\rm T}$. This means that finite $\Delta_{\rm ic}$ only brings a further deviation from the experimental magnetization. This confirms that the simple AFM Ising interchain coupling gives the best fit.

\subsection{$z$-component of local spin}

\begin{figure}[t]
\centering
\includegraphics[width=0.9\linewidth]{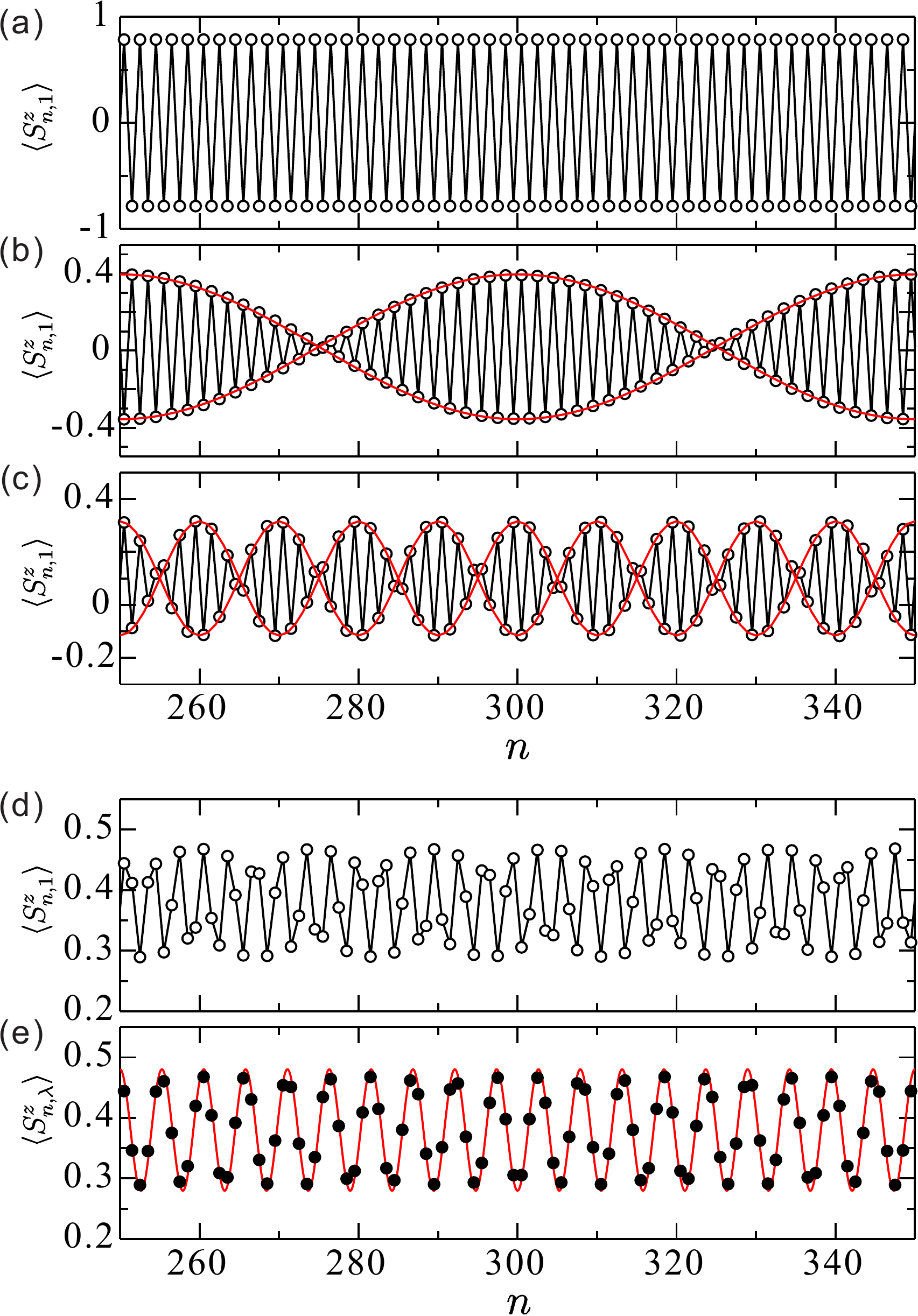}
\caption{
$z$-component of local spin $\langle S^z_{i,1}\rangle$ for (a) $H=0-0.35$~T, (b) 0.36~T, (c) 0.44~T, and (d) 0.89~T. (e) $\langle S^z_{n,\lambda}\rangle$ with $\lambda=\frac{3+(-1)^n}{2}$ for $H=0.89$~T (see text). The red line is a fitting by Eq.~\eqref{Szgeneral}.
}
\label{localSz}
\end{figure}

Experimentally, an AFM-IC phase transition has been observed at $H=0.35$~T~\cite{Wu19}. To investigate the nature of the AFM-IC transition and the stability of the IC ordering, we calculate the $z$-component of local spin, $\langle S^z_{n,j}\rangle$. Here, open boundary conditions are applied. This allows us to directly observe a translation-symmetry broken state due to Friedel oscillations. In Fig.~\ref{localSz}(a-d) we show $\langle S^z_{n,j}\rangle$ as a function of the position $n$ in either of the chains for several strengths of the magnetic field. We use an open cluster with $600\times2$ sites for the system (\ref{ham1}) and the result for the central $100$ sites is plotted. Since the system \eqref{ham1} is unfrustrated, by considering the Fermi wavenumber after Jordan-Wigner transformation, a possible form of $\langle S^z_{n,j}\rangle$ is derived as
\begin{align}
\langle S^z_{n,j}\rangle=\pm A_L(-1)^n\cos(2\pi Mn+\phi)+M,
\label{Szgeneral}
\end{align}
where $A_L$ is the amplitude of oscillation and the $\pm$ sign refers to the two sublattices ($j=1$ or $2$).

At $H=0$, Eq.~\eqref{Szgeneral} leads to $\langle S^z_{n,1}\rangle=\pm A_L(-1)^n$. As shown in Fig.~\ref{localSz}(a), this is a pure AFM ordering and remains unchanged up to the critical field $H=0.35$~T. In consistency with the experimental observation, the IC modulation appears at $H>0.35$~T. At $H=0.36$~T, which is slightly higher than the critical field, an IC modulation with long wavelength is clearly seen [Fig.~\ref{localSz}(b)]. By fitting the modulation with Eq.~\eqref{Szgeneral}, we estimate the wavelength $\Delta n \sim 100$ ($M \sim 0.01$) and $A_{600}=0.376$. With increasing field, $\Delta n$ becomes shorter and $A_L$ decreases. At $H=0.44$ T [Fig.~\ref{localSz}(c)], we estimate $\Delta n \sim 20$ ($M \sim 0.05$) and $A_{600}=0.216$. In the low-field region, the fitting with Eq.~\eqref{Szgeneral} is rather straightforward because the wavelength is much longer than the lattice spacing, i.e., $\Delta n \gg 1$. However, if the wavelength is comparable with the lattice spacing, i.e., $\Delta n \sim {\cal O}(1)$, the fitting is not very simple. In Fig.~\ref{localSz}(d) the profile of  $\langle S^z_{n,1}\rangle$ at $H=0.89$ T is shown. It  looks unclear how to fit it with Eq.~\eqref{Szgeneral}. This problem can be solved by plotting, for example, $\langle S^z_{n,1}\rangle$ for odd $n$ and $\langle S^z_{n,2}\rangle$ for even $n$, namely, $\langle S^z_{n,\lambda}\rangle$ with $\lambda=\frac{3+(-1)^n}{2}$. Fig.~\ref{localSz}(e) shows $\langle S^z_{n,\lambda}\rangle$ at $H=0.89$ T. In this way, the fitting with Eq.~\eqref{Szgeneral} can be handily performed and we estimate $\Delta n \sim 5$ ($M \sim 0.2$). This method works in the whole range of IC phase.

\begin{figure}[t]
\centering
\includegraphics[width=1.00\linewidth]{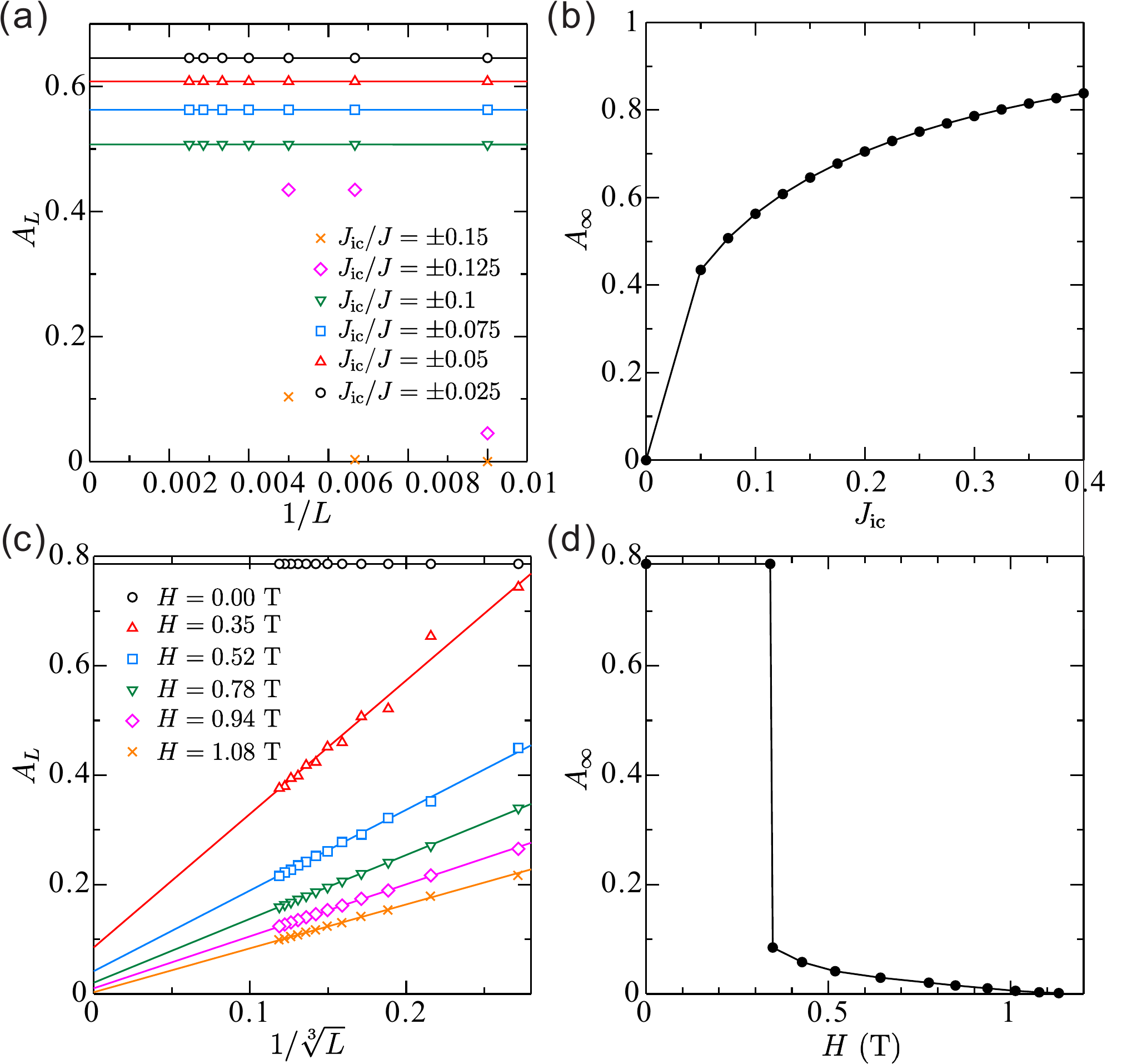}
\caption{
(a) Finite-size scaling analysis for the amplitude of AFM modulation at $M=0$ for several strengths of the Ising interchain coupling $J_{\rm ic}$. (b) $L\to\infty$ extrapolated values of the amplitude as a function of $J_{\rm ic}$. (c) Finite-size scaling analysis for the amplitude of IC oscillation for several strengths of magnetic field $H$ with our YbAlO$_3$ parameters ($J=2.3$~K, $J_{\rm ic}=0.8$~K). (b) $L\to\infty$ extrapolated values of the amplitude as a function of $H$ for YbAlO$_3$.
}
\label{ICscaling}
\end{figure}

It is important to examine whether the AFM and IC modulations are long-range ordered. To determine it, we perform finite-size scaling analysis of the amplitude $A_L$. We look first at the stability of the AFM order as a function of the Ising interchain coupling. In Fig.~\ref{ICscaling}(a) the scaling of $A_L$ in the AFM phase, i.e., at $M=0$, is shown for several strengths of the interchain coupling. Note that the FM and AFM Ising couplings give exactly the same results in this case. The finite-size effect seems to be very small once the system size is extended beyond the critical length. The extrapolated values of $A_L$ in the thermodynamic limit $L \to \infty$ are plotted as a function of $J_{\rm ic}$ in Fig.~\ref{ICscaling}(b). We can confirm that the AFM long-range order is stabilized only if the chains are coupled through Ising interaction. In the AFM phase ($0<H<0.35$T) our YbAlO$_3$ parameter $J_{\rm ic}/J=0.35$ leads to $A_\infty=0.393$, which is slightly reduced from the full value $0.5$ by the intrachain quantum fluctuations.

Next, we turn to the IC state, i.e., at $0<M<0.5$. In Fig.~\ref{ICscaling}(c) the finite-size scaling is shown for several strengths of magnetic field. For all $H$, the best scaling function seems to be linear by choosing the horizontal axis as $1/\sqrt[3]{L}$. This might be related to the three-dimensional ordering of the IC modulation. Fig.~\ref{ICscaling}(b) shows the extrapolated values of $A_\infty$ as a function of magnetic field. At the critical field $H=0.35$~T, $A_\infty$ drops down from 0.393 to 0.042, indicating a first order transition between the AFM and IC phases transition. With further increasing magnetic field, the value of $A_\infty$ decreases gradually and becomes zero at the saturation field $H=1.21$~T, indicating a second order or continuous transition. Thus, we find that $A_\infty$ is small but finite at $0.35~{\rm T}<H<1.21~{\rm T}$. This means that the IC modulation is long-range ordered in the whole range of $0<M<0.5$. To support this statement, we consider the central charge in the next subsection.

\subsection{Central charge}

\begin{figure}[t]
\centering
\includegraphics[width=1.00\linewidth]{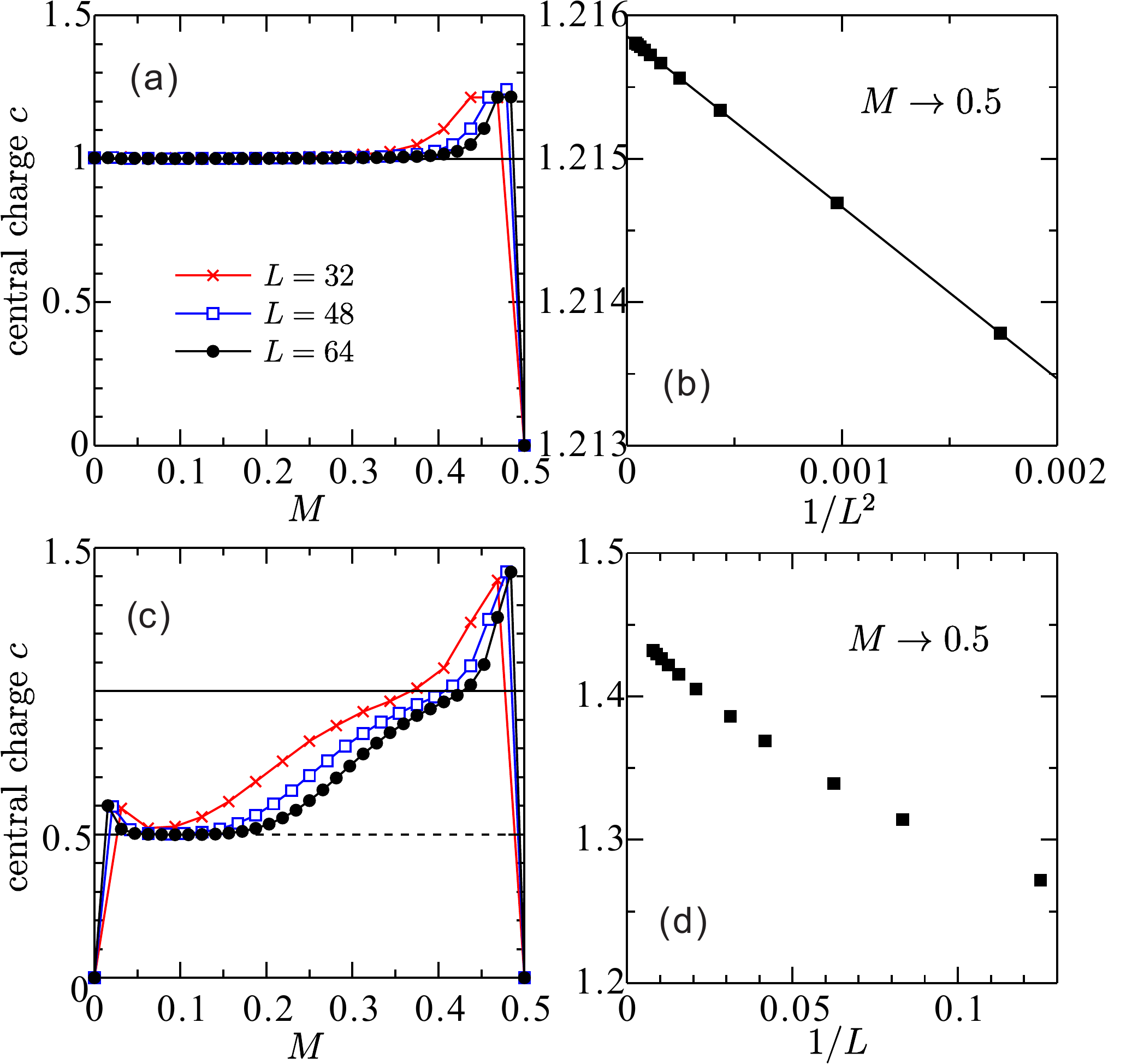}
\caption{
Central charge for (a) the isotropic SU(2) Heisenberg chain and (c) the system~\eqref{ham1} as a function of the magnetization. (b)(d) Finite-size scaling analysis of the central charge to the saturation limit $M\to0.5$. The line in (b) is a linear fit.
}
\label{central_charge}
\end{figure}

In general, the saturation field at $T=0$ is a quantum critical point. In YbAlO$_3$, the TLL behavior has been experimentally observed near the saturation field. To investigate the consistency with our model \eqref{ham1}, we calculate the central charge $c$ which provides definitive information on the universality class of $(1+1)$ dimensional system~\cite{Cardy96}. A system in the TLL phase belongs to the Gaussian universality class ($c=1$) and $c<1$ is expected for the gapped phase from the renormalization in the massive region. The central charge can be numerically calculated through the von Neumann entanglement entropy $S_L(l)=-{\rm Tr}_l \rho_l \log \rho_l$, where $\rho_l={\rm Tr}_{L-l}\rho$ is the reduced density matrix of the subsystem with length $l$ and $\rho$ is the full density matrix of the whole system with length $L$. Using  conformal field theory (CFT), the relation between $S_L(l)$ and $c$ has been derived~\cite{Affleck91,Holzhey94,Calabrese04}: $S_L(l)=\frac{c}{3}\ln\left[\frac{L}{\pi}\sin\left(\frac{\pi l}{L}\right)\right]+s_1$, where $s_1$ is a non-universal constant. A prime objective of using this formula is to estimate the central charge~\cite{Laflorencie06,Legeza07}. For the system \eqref{ham1} under periodic boundary conditions, the central charge is obtained via~\cite{cc}
\begin{align}
N_{\rm c}=\frac{3\left[S_L\left(\frac{L}{2}-1\right)-S_L\left(\frac{L}{2}\right)\right]}
{\ln\left[\cos\left(\frac{\pi}{L}\right)\right]},
\label{eq:centralcharge}
\end{align}
where $N_{\rm c}$ is the number of chains considered.

For reference, let us first look at the central charge for an isotropic SU(2) Heisenberg chain. It is plotted as a function of the magnetization $M(=S^z_{\rm tot}/(N_{\rm c}L))$ for several system lengths $L$ in Fig.~\ref{central_charge}(a). As expected, the system is always critical $c=1$ at $0 \le M<0.5$. Near the saturation $M=0.5$, it goes up steeply. In the thermodynamic limit, there would be a jump at $M=0.5$, which signifies the quantum criticality at the saturation field. In Fig.~\ref{central_charge}(b) a finite-size scaling analysis of the central charge with keeping $S^z_{\rm tot}=\frac{L-2}{2}$ ($M=\frac{L-2}{2L}$) as a function of $1/L^2$ is performed. The thermodynamic limit $L\to\infty$ corresponds to the saturation limit $M\to0.5$. Thus, we obtain $c=1.215854$ in the limit of $M=0.5$. The origin of this $c$ value is still unknown and it should be further studied in the future.

Then, we turn to our model \eqref{ham1}. In Fig.~\ref{central_charge}(c) the central charge of the system~\eqref{ham1} is plotted as a function of the magnetization for several system lengths. At $M=0$, we find $c=0$ as expected from the gapped ground state. At $0<M<0.5$, interestingly unlike in the case of isotropic SU(2) Heisenberg chain, the central charge approaches the Ising universality class ($c=0.5$) with increasing $L$, as the SU(2) symmetry is reduced to U(1) symmetry due to the Ising interchain coupling. Near the saturation $M=0.5$, the convergence to $c=0.5$ seems to be very slow with $L$ since the system is nearly critical with the friable IC order. Nevertheless, the crossing point with $c=1$ for $L=64$ is already $M\sim0.43$ close to the saturation $M=0.5$, namely, $c<1$ at $0<M<0.43$. Therefore, it is most likely that the system belongs to the Ising universality class over the whole region of $0<M<0.5$ in the thermodynamic limit. The central charge jumps from $c=0.5$ to $\approx1.5$ at $M=0.5$. CFT~\cite{Popkov05,Olalla11} predicts the value of $c=1.5$ for the ferromagnetic point. In Fig.~\ref{central_charge}(d) the central charge at fixed $M=(L-1)/(2L)$  for a system with length $L$ is plotted as a function of $1/L$. As $1/L\to0$, i.e., $M$ approaches the saturation value $M=0.5$, the central charge approaches $1.5$. Therefore, we can confirm that the saturation field is a quantum critical point. This is consistent with experimental observations~\cite{Wu19}.

\subsection{Dynamical spin structure factor}

\begin{figure*}[t]
\centering
\includegraphics[width=0.75\linewidth]{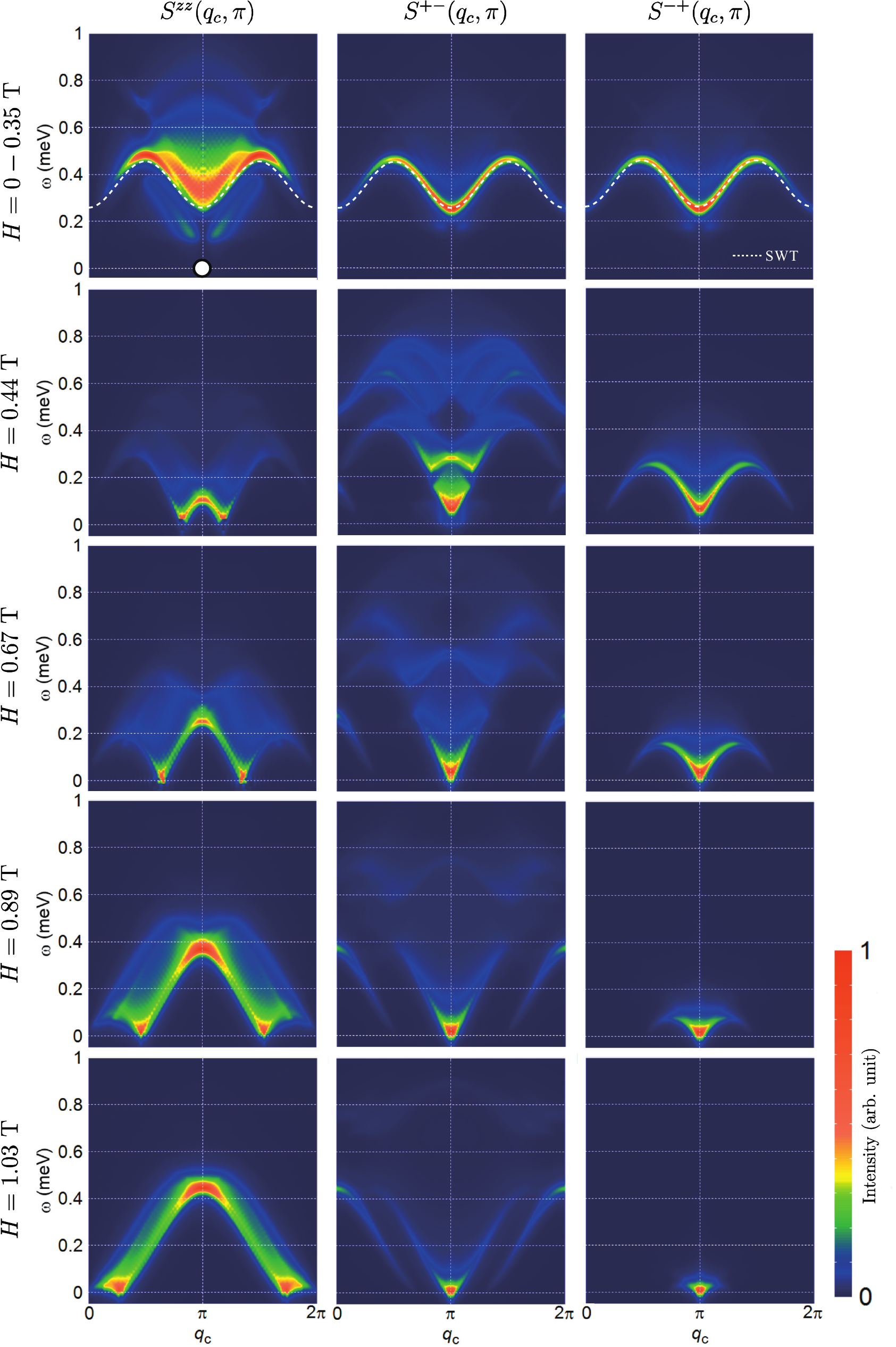}
\caption{
(a) Dynamical spin structure factors $S^{zz}(q_c,\omega)$ (left panel), $S^{+-}(q_c,\omega)$ (center panel), and $S^{-+}(q_c,\omega)$ (right panel) for several strengths of the magnetic field $H \parallel z$. Open circle in the left-top panel denote a large $\delta$-peak indicating the AFM order. The dotted lines in the top panels are magnon dispersion obtained within the spin-wave analysis.
}
\label{fig_Sqw}
\end{figure*}

For calculating dynamical properties, we use the dynamical DMRG method. To examine the low-energy excitations and their development with the magnetic field, we calculate the dynamical spin structure factor, defined as
\begin{eqnarray}
\nonumber
S^{\gamma\bar{\gamma}}(\vec{q},\omega) &=& \frac{1}{\pi}{\rm Im} \langle \psi_0 | (S^\gamma_{\vec{q}})^\dagger \frac{1}{\hat{H}+\omega-E_0-{\rm i}\eta} S^\gamma_{\vec{q}} | \psi_0 \rangle\\
&=& \sum_\nu |\langle \psi_\nu |S^\gamma_{\vec{q}}| \psi_0 \rangle|^2 \delta(\omega-E_\nu+E_0),
\label{spec}
\end{eqnarray}
where $\gamma$ is $z$ or $-(+)$, $| \psi_\nu \rangle$ and $E_\nu$ are the $\nu$-th eingenstate and the eigenenergy  of the system, respectively ($\nu=0$ corresponds to the ground state). Under open boundary conditions, we define the momentum-dependent spin operators as
\begin{align}
S^\gamma_{\vec{q}} = \sqrt{\frac{2}{L+1}} \sum_l e^{i{\vec{q}}\cdot{\vec{r}}} S^\gamma_{\vec{r}},
\label{operator_pbc}
\end{align}
with (quasi-)momentum $\vec{q}=(\pi Z_x/(L+1), \pi Z_y)$ for integers $1 \le Z_x \le L$ and $Z_y=0, 1$. We use an open cluster with $50\times2$ sites. Due to the AFM correlation between the neighboring chains, we restrict ourselves to the case $\vec{q}=(q_c,\pi)$ where $q_c$ is momentum along the chain ($c$-axis). In Fig.~\ref{fig_Sqw} DMRG results of the dynamical structure factors are shown for several strengths of magnetic field. The left, middle, and right panels corresponds to $S^{zz}(q_c,\omega)$, $S^{+-}(q_c,\omega)$, and $S^{-+}(q_c,\omega)$, respectively.

At low magnetic fields ($H=0-0.35$~T), the three spectra show an explicit gap $\sim0.22$~meV at $q_c=\pi$. The longitudinal structure factor $S^{zz}(q_c,\omega)$ has a dominant $\delta$-peak at $(q_c,\omega)=(\pi,0)$ indicating the AFM order. The weight of this peak is about 90$\%$ of the total weight $\int S^{zz}(q_c=\pi,\omega)d\omega$ and it is denoted by an open circle in Fig.~\ref{fig_Sqw}. The thick (or blur) dispersion of $S^{zz}(q_c,\omega)$ is a typical signature of Ising-type spin anisotropy, which is effectively induced by the Ising interchain coupling. The two-spinon continuum is seen but seems to be somewhat suppressed in comparison to those of the SU(2) Heisenberg chain. These features are clearly visible in the experimental spectra [see Ref.~\onlinecite{Wu19}]. In the AFM phase, $S^{+-}(q_c,\omega)$ and $S^{-+}(q_c,\omega)$ are equivalent: Their dispersions are very thin and the two-spinon continuum is further dilute. Since the AFM order is long ranged and stable, a  magnon excitation picture may be a good approximation. In fact, the lower bound of the spectra is well described by the spin wave analysis $\omega(q_c)=J^\prime+J\sin^2q_c$, which is shown with white dotted lines in Fig.~\ref{fig_Sqw}.

As soon as the system goes into the IC phase, the spectra are drastically changed as a consequence of the first-order transition. Let us see $S^{zz}(q_c,\omega)$. At $H>0.35$~T, the single large peak at $(q_c,\omega)=(\pi,0)$ in the AFM phase is split into two peaks at $(q_c,\omega)=((1\pm2M)\pi,0)$; thus, $q_c$ is shifted from $\pi$ to $0$, $ 2\pi$ with increasing the magnetic field. This $q_c$ value corresponds to the propagation number of the IC long-range order. The magnetization $M$ is $0.09$, $0.18$, $0.27$, and $0.36$ for $H=0.44$, $0.67$, $0.89$, and $1.03$~T, respectively, and lead to $q_c=(1\pm0.18)\pi$, $(1\pm0.36)\pi$, $(1\pm0.54)\pi$, and  $(1\pm0.72)\pi$. They perfectly agree with our DMRG spectra. Remarkably, the total weight of $\int S^{zz}(q_c=\pi,\omega)d\omega$ seems to remain comparable to that of the IC peaks even at high field, although the IC order must be dominant. This weight comes from the staggered oscillation of local spin $\langle S^z_i \rangle$ induced by the Ising interchain coupling [see Fig.~\ref{localSz}]. Therefore, the IC state could also be interpreted as the coexistence of two spin-density waves with $q_c=\pi$ and $(1\pm2M)\pi$. Similarly to the AFM phase, the two-spinon continuum seems to be weakened due to the stability of the long-range order. Our DMRG spectra in the IC phase also agree with the experimental ones reported in Ref.~\onlinecite{Wu19}.

We then turn to the transverse structure factors $S^{+-}(q_c,\omega)$ and $S^{-+}(q_c,\omega)$. Their low-energy part can be interpreted by considering those of a single Heisenberg chain as the $x$- and $y$-component of spins are uncoupled between chains. It is known that AFM $S^x_iS^x_j$ and  $S^x_iS^x_j$ correlations grow with increasing the magnetic field in the Heisenberg chain~\cite{Affleck99}. Accordingly, a large peak appears at $(q_c,\omega)\approx(\pi,0)$ in the whole region of the IC phase. The magnon dispersion (lower bound of the continuum) is explained by that of the XY chain under magnetic field along the $z$-axis. The dispersion of $S^{-+}(q_c,\omega)$ is a sine-like function with nodes at $q_c=\pi$, $2\pi M$, and $2\pi(1-M)$ and width $J[1-\sin{\frac{\pi}{2}(1-2M)}]$; that of $S^{+-}(q_c,\omega)$ is a sine-like function with nodes at $q_c=\pi$, $2\pi(1+M)=2\pi M$, and $-2\pi M=2\pi(1-M)$ and width $J[1+\sin{\frac{\pi}{2}(1-2M)}]$. Furthermore, it is worth noting that in the higher energy range around $\omega\sim0.5-0.8$~meV there exist another continuum in $S^{+-}(q_c,\omega)$. They are nearly separated from the low-energy structures by $\sim J^\prime$; and thus, reveal correlations between the two sublattices. If the transverse structure factors could be experimentally observed, more sophisticated study on YbAlO$_3$ would be allowed.


\subsection{Effective 1D model: Heisenberg chain with staggered field}

The effect of the Ising interchain coupling $S^z_{i,j} S^z_{i,j^\prime}$ may be mimicked by a self-consistent staggered magnetic field $(-1)^iS^z_i$~\cite{Schulz96,Bocquet01}. If this is the case, our model \eqref{ham1} could be mapped onto a single Heisenberg chain with a staggered magnetic field:
\begin{align}
	H = J \sum_i \vec{S}_i\cdot\vec{S}_{i+1}+ (-1)^ih_zS^z_i.
	\label{ham2}
\end{align}
In fact, this model was used to analyze some experimental observations in Ref.~\onlinecite{Wu19}. A simplest mapping of the model \eqref{ham1} onto the model \eqref{ham2} can be performed by comparing the low-lying energies of a small cluster. We take a 4-site plaquette for \eqref{ham1} and a 2-site dimer for \eqref{ham2}: The ground state energies in the $S^z=0$ sector are $-\frac{J}{2}-\sqrt{J^2+\frac{J^{\prime2}}{4}}$ and $-\frac{J}{4}-\sqrt{\frac{J^2}{4}+h^2}$; and those in the $S^z=1$ sector are $-\frac{J}{2}$ and  $-\frac{J}{4}+\sqrt{\frac{J^2}{4}+h^2}$, respectively. By comparing the energy differences between two spin sectors, i.e., $\sqrt{J^2+\frac{J^{\prime2}}{4}}=2\sqrt{\frac{J^2}{4}+h^2}$, it leads to $h=\pm\frac{J^\prime}{4}$. Thus, we estimated the parameters for the effective single spin chain as $J=2.3K$ and $h=0.8K$. These values agrees well to those ($J=2.4$~K, $h=0.66$~K used in Ref.~\onlinecite{Wu19}.

\section{Conclusion}

Using the DMRG technique we studied isotropic AFM Heisenberg chains coupled by AFM Ising interaction as an effective spin model for the ytterbium aluminum perovskite YbAlO$_3$. The effective intrachain and interchain couplings were estimated as $J=2.3$~K and $J_{\rm ic}=0.8$~K, respectively, from the fitting of the experimental magnetization curve. At $0.35~{\rm T}<H<1.21~{\rm T}$ ($0<M<0.5$), a long-ranged IC order is stabilized by the Ising interchain coupling; accordingly, the system belongs to the Ising universality class ($c=0.5$) in the IC phase. At the saturation field, the central charge jumps from $c=0.5$ to a FM value $c=1.5$ which is a signature of QCP. Our results fully agree with the experimental observations reported in Ref.~\onlinecite{Wu19}. Furthermore, we calculated the dynamical structure factors. Our calculations quantitatively reproduce the low-energy excitations in the experimental spectra. In this paper, we focused on the case of $H \parallel a$, as the low-temperature experimental data are available only for this case. For more sophisticated theoretical analysis, measurements for other field directions are required.

\section*{Acknowledgements}

This work is supported by the DFG through the W\"urzburg-Dresden Cluster of Excellence on Complexity and Topology in Quantum Matter -- \textit{ct.qmat} (EXC 2147, project-id 39085490) and through SFB 1143 (project-id 247310070). We thank S. E. Nikitin for fruitful discussion and the experimental magnetization data and U. Nitzsche for technical assistance.

\appendix

\section{Intrachain XXZ anisotropy}

\begin{figure}[t]
\centering
\includegraphics[width=0.6\linewidth]{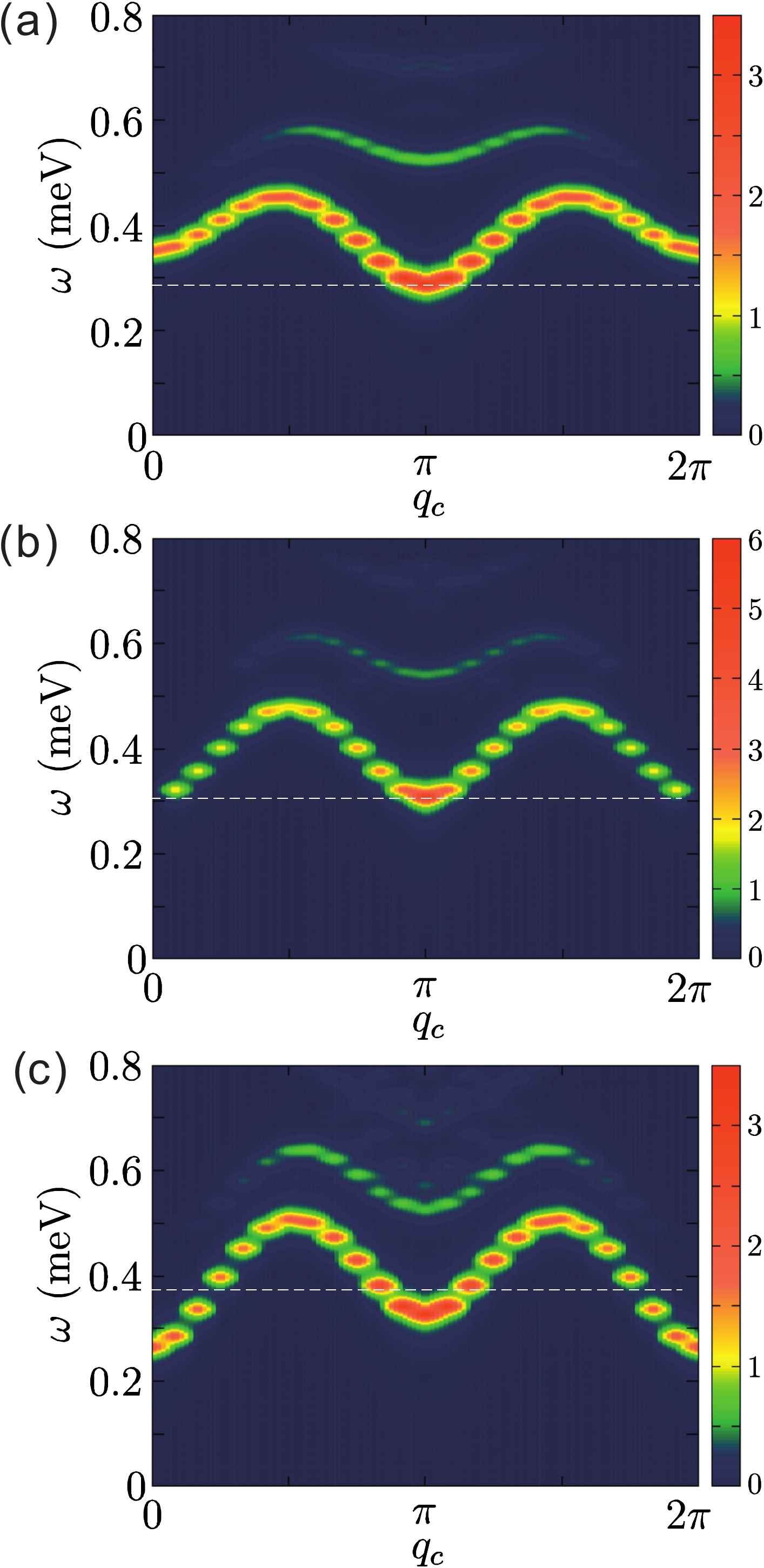}
\caption{
Dynamical spin structure factor for the spin chain \eqref{ham3} with (a) $\Delta=1.43$, (b) $\Delta=1$, and (c) $\Delta=0.77$ using a periodic 24-site cluster.
}
\label{spec_XXZ}
\end{figure}

Generally, the exchange interaction of rare earth ions has a typical form $\vec{S}\cdot\vec{S}$. Thus, despite the highly anisotropic doublet state of Yb$^{3+}$ ions, we assumed that the exchange coupling along the chain direction is isotropic, i.e., an XXX chain in the main text. To verify this assumption, we consider the XXZ-anisotropy dependence of the dynamical spin structure factor. We then introduce the XXZ anisotropy into the effective single chain \eqref{ham2} as
\begin{align}
	H = J \sum_i \frac{\Delta}{2}(S^+_iS^-_{i+1}+S^-_iS^+_{i+1})+S^z_iS^z_{i+1} + (-1)^ih_zS^z_i.
	\label{ham3}
\end{align}
In Fig.~\ref{spec_XXZ} we show the result for the dynamical spin structure factor for several $\Delta$ values with a periodic 24-site cluster. In the experimental spectra, the lowest excitations, i.e., gaps, at $q_c=0$ and $q_c=\pi$ are nearly the same for the AFM state. As seen in Fig.~\ref{spec_XXZ}, this situation is realized at $\Delta\sim1$. Therefore, it is reasonable to assume the intrachain interaction to be isotropic.

\section{Magnetization with ferromagnetic Ising interchain coupling}

\begin{figure}[tbh]
\centering
\includegraphics[width=0.7\linewidth]{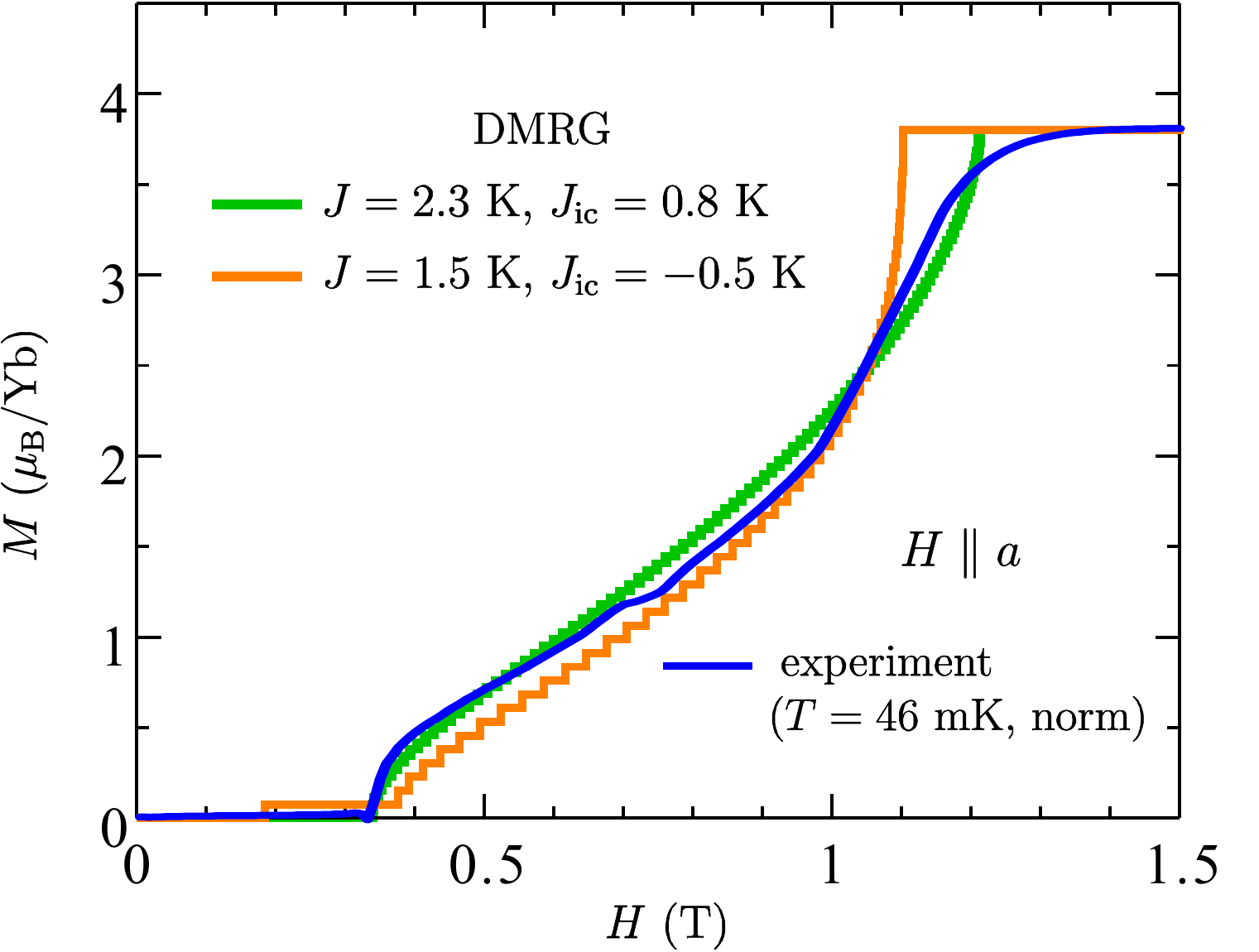}
\caption{
Fitting of the low-temperature experimental magnetization curve for $H \parallel a$ by DMRG result with our effective spin model \eqref{ham1}, where $g^z=g^a=7.6$ is used.
}
\label{mag_fitting_Appendix}
\end{figure}

As shown in the main text, the best fit of the experimental magnetization was achieved by taking the Ising interchain coupling to be AFM in our effective spin model~\eqref{ham1}. If we assume it to be FM, the fitting is less accurate (see Fig.~\ref{mag_fitting_Appendix}): In the case of FM Ising interchain coupling, (i) a sharp rise of the experimental $M$ at the AFM-IC phase transition $H=0.35$~T can not be reproduced, and (ii) the increase of $M$ around the saturation field is too steep in comparison to the experimental curve. Therefore, we decided to focus on the case of AFM Ising interchain coupling in the main text. If the low-temperature magnetization curves for the other field directions are made available, more sophisticated fitting analysis may be performed.

\section{Static structure factor}

\begin{figure}[tbh]
\centering
\includegraphics[width=0.8\linewidth]{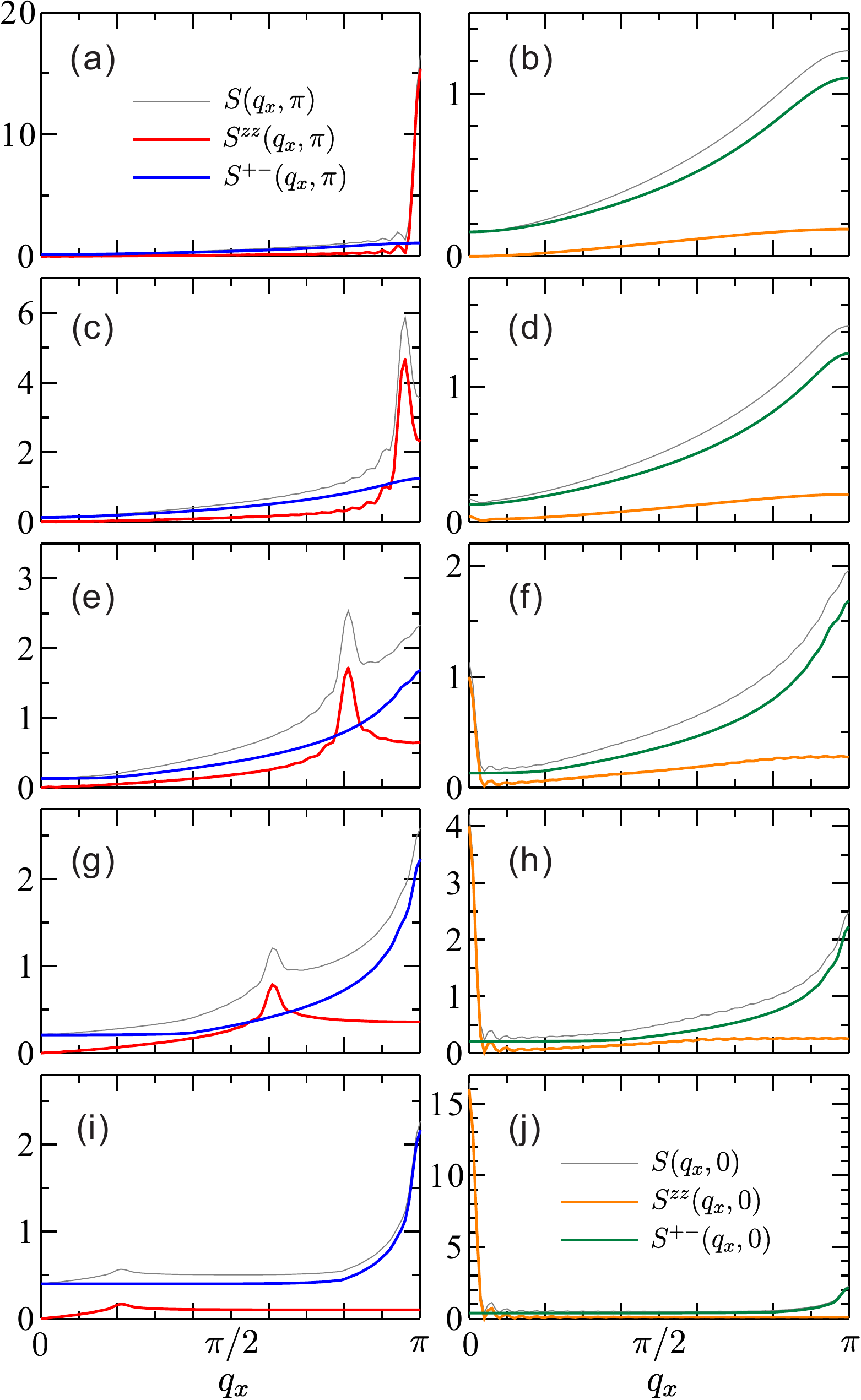}
\caption{
Static spin structure factors for (a,b) $H=0-0.35$~T, (c,d) $H=0.44$~T, (e,f) $H=0.67$~T, (g,h) $H=0.89$~T, and (i,j) $H=1.03$~T using an open cluster with $50\times2$ sites.
}
\label{fig_Sq}
\end{figure}

To see the magnetic structure in the ground state, we calculate the static spin structure factors. They are defined by a Fourier transform of the spin-spin correlation functions
\begin{align}
S^{\gamma\bar{\gamma}}(\vec{q}) = \sum_{i,j} e^{i\vec{q}\cdot(\vec{R}_j-\vec{R_i})} \langle S^\gamma S^{\bar{\gamma}} \rangle.
\end{align}
\label{Sq}
We plot the results of static spin structure factors in Fig.~\ref{fig_Sq}. For $H=0-0.35$~T, the longitudinal structure factor $S^{zz}(\vec{q})$ has a large peak at $\vec{q}=(\pi,\pi)$ indicating the AFM order. As soon as the system goes into the IC phase, this peak is abruptly lowered and the peak position is shifted to small $q_c$ with increasing the magnetic field. On the other hand, a FM peak at $\vec{q}=(0,0)$ in $S^{zz}(\vec{q})$ begins to grow around $H=0.67$~T. The transverse structure factor $S^{+-}(\vec{q})$ always has a peak at $\vec{q}=(\pi,0)$ and $\vec{q}=(\pi,\pi)$; the peak becomes sharper with increasing the magnetic field because of the enhancement of AFM $S^x_iS^x_j$ and $S^x_iS^x_j$ correlations.

\bibliography{YbAlO3}

\end{document}